\documentclass[rnote, traditabstract]{aa} 
\usepackage{subfigure}                                   
\usepackage{epsfig}
\usepackage{wasysym}
\usepackage{graphicx}
\usepackage{natbib}
\usepackage{lscape}
\usepackage{multirow}

\usepackage{color}
\providecommand{\tabularnewline}{\\}
\usepackage{txfonts}
\begin{document}

\title{Searching for solar siblings among the HARPS data\thanks{Based on observations collected at the La Silla Paranal Observatory,
ESO (Chile) with the HARPS spectrograph at the 3.6-m telescope (ESO
runs ID 72.C-0488, 082.C-0212, and 085.C-0063).}}


\titlerunning{Tc slope}


\author{S.~F.~A.~Batista\inst{1} 
\and V.~Zh.~Adibekyan \inst{1}
\and S.~G.~Sousa\inst{1,2,3}
\and  N.~C.~Santos\inst{1,2}
\and  E.~Delgado~Mena\inst{1}
\and A.~A.~Hakobyan\inst{4}
}

\institute{Centro de Astrof\'{\i}ísica da Universidade do Porto, Rua das Estrelas,
4150-762 Porto, Portugal\\
\email{sergio.batista@astro.up.pt}
\and Departamento de F\'{\i}ísica e Astronomia, Faculdade de Ci\^{e}ncias da Universidade do Porto, Portugal
\and Instituto de Astrof\'{\i}sica de Canarias, 38200 La Laguna, Tenerife, Spain
\and Byurakan Astrophysical Observatory, 0213 Byurakan, Aragatsotn province, Armenia}

   \date{Received ... / Accepted ...}

 
\abstract
{The search for the solar siblings has been particularly fruitful in the last few years. 
Until now, there are four plausible candidates pointed out in the literature: HIP21158, HIP87382, HIP47399, and HIP92831. 
In this study we conduct a search for solar siblings among the HARPS high-resolution FGK dwarfs sample, which 
includes precise chemical abundances and kinematics for 1111 stars. Using a new approach based on chemical 
abundance trends with the condensation temperature, kinematics, and ages we found one (additional) potential solar sibling candidate: HIP97507.}


\maketitle
%

\section{Introduction}

Nowadays, it is accepted that most stars are born in clusters \citep{Lada-03}, as it is also believed to have happened in the case of our Sun. 
According to \cite{Wielen-1996}, the Sun was born in the inner part of the Milky Way, perhaps 1.9 kpc closer to the galactic center than its 
current location. The stars that have born in the same cluster as the Sun, called solar siblings, were spread away from their initial orbits 
over the Galaxy. In this scenario, mechanisms such as spiral density waves may play an important role \citep[e.g.][]{Lepine-03}. 

\cite{Zwart-09} quote that if the parental cluster of our Sun had 10$^{3}$ stars, at least 1\% of the solar siblings (about 10 - 60) should 
still be located within 100 pc and that more than 10\% should be within a region of 1 kpc, around the present location of our Sun. On their hand,  
\cite{Mishurov-11} showed that this result is only true in the most optimistic case.

Identifying the Sun's siblings would put constraints on the number of stars in the original cluster. Moreover, reconstructing 
the orbits of the siblings in the Galaxy would help us to understand the Sun's accurate birth location.

The search for solar siblings is an ambitious task, strongly restricted by their ages, metallicities, and kinematics. 
The latter has been particularly fruitful during the last few years. A search for solar siblings conducted among the Hipparcos catalogue, 
provided a list of 6 potential candidates \citep{Brown-10}: HIP21158; HIP30344; HIP51581; HIP80124; HIP90122; HIP99689. Due to relatively low age 
estimates or low metallicity values or high radial velocities, only the star HIP21158 was pointed out as being a real potential candidate. 
The same work, also discussed five more stars, which were found close to the solar isochrones: HIP56287, HIP57791, HIP89825, HIP92381, 
and HIP101911. However, all of these stars were discarded either because of their radial velocities, or due to the lack of age 
estimates or metallicity values. \cite{Bobylev-11} introduced a new kinematic approach to search for solar siblings. 
They found two potential candidates: HIP87382 and HIP47399. More recently, the first giant, HIP175740, was
pointed out as a potential solar sibling candidate \citep{Batista-12}. In addition, they also discuss a potential candidate, which has 
known orbiting planets (HIP115100). However, the latter star was discarded due to its supersolar metallicity.

Very recently, \cite{Adibekyan-14} used a sample of 148 solar-type stars from \cite{Jonay-10, Jonay-13} to explore the main factors 
responsible for the abundance trends with condensation temperature (T$_{c}$ hereafter). The authors found that the slope of this trend 
(T$_{c}$ slope hereafter)  significantly correlates (at more than 4$\sigma$) with the stellar age. 
They also found an evidence that the T$_{c}$ slope correlates with the mean galactocentric distance of 
the stars (R$_{mean}$), indicating that stars originated in the inner Galaxy have less refractory elements relative to the volatiles.
Thus the authors concluded that the age and the Galactic birth place are determinant for the chemical
structure of stars and they also determine the T$_{c}$ slopes. 
This means that stars which were born in the same cluster should have a similar metallicity and 
should show similar abundance trends with the  T$_{c}$.

The search for solar siblings is strongly dependent on the ability of finding compatible metallicities between surveys \citep{Batista-12}. 
In this work, we conducted a search for solar siblings among the HARPS high-resolution FGK dwarfs sample \citep{Adibekyan-12a} using a new
approach based on the observed chemical abundance trends  with the  T$_{c}$.
The paper is organized as follows. In Sect. 2, we present the sample. Section 3 presents the considered criteria to conduct this search 
and the potential candidates. Finally, in Sect. 4, we summarize the results.

\section{The sample}

Our initial sample comprises 1111 FGK dwarfs with high resolution spectra observed with the HARPS spectrograph \citep{Mayor-03} at the ESO 3.6-m 
telescope (La Silla, Chile). The sample was built to study the chemical properties of exoplanet hosting stars 
\citep[e.g.][]{Adibekyan-12b,Adibekyan-12c} and then used to study chemical properties and kinematics of stellar populations in the solar 
neighborhood \citep[e.g.][]{Adibekyan-11,Adibekyan-13a}.

Precise stellar parameters (T$_{eff}$, $\log\,g$, [Fe/H], and $\xi_{t}$) and elemental abundances for 12 elements (namely Na, Mg, Al, Si, Ca, Ti, Cr, Ni, Co, 
Sc, Mn, and V) for all the stars were determined in a homogeneous way. The atmospheric parameters and elemental abundances were determined 
using a local thermodynamic equilibrium analysis relative to the Sun with the 2010 revised version of the spectral synthesis 
code MOOG \citep{Sneden-73} and a grid of Kurucz ATLAS9 plane-parallel model atmospheres \citep{Kurucz-93}. The reference abundances used 
in the abundance analysis were taken from \cite{Anders-89}. 
Since all the analysis is differential with respect to the Sun, the reference abundances are not a major concern.
We refer the reader to \cite{Sousa-08}  and \cite{Adibekyan-12a} for more details.

The stars in the initial sample had effective temperatures ranging between 4487 and 7212 K. However only a few stars in the sample have temperatures 
that are very different from the solar one. Recently, \cite{Tsantaki-13} showed that
the parameters of the stars in this sample are most precise if the T$_{eff}$ is higher than 5000 K. 
To have the most reliable stellar parameters and abundances \citep[see also][]{Adibekyan-13a} we decided to establish a
cutoff in temperature at 5000 K. This lead our sample to a 881 FGK stars (mostly dwarfs, with only 23 having $\log\,g$ $<$ 4 dex). 
The metallicites of the stars range  from -1.39 to 0.55 dex.

For all the stars in the sample, \cite{Adibekyan-12a} derived the space velocity components (U, V, W) with respect to the 
Local Standard of Rest, adopting the standard solar motion 
(U$_{\odot}$, V$_{\odot}$, W$_{\odot}$) = (11.1, 12.24, 7.25) km $\mathrm{s}^{-1}$ from \citet{Schonrich-10}.

\subsection{The T$_{c}$ slope derivation}

\begin{figure}
\begin{center}
\begin{tabular}{c}
\includegraphics[angle=270,width=0.9\linewidth]{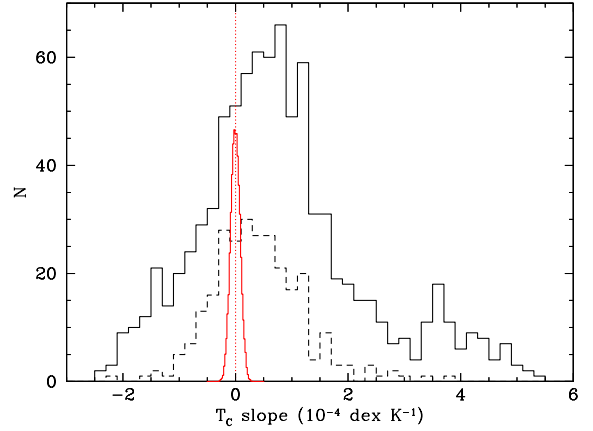}
\end{tabular}
\end{center}
\vspace{-0.6cm}
\caption{The distribution of the  T$_{c}$ slopes for the full sample (black solid line) and for the stars with $\mid$[Fe/H]$\mid$ $<$ 0.1 dex 
(black dashed line). The position of the Sun is shown by red dotted line. The expected distribution of the T$_{c}$ slopes for 500 solar composition stars 
accepting the average abundance errors from \cite{Adibekyan-12a} is plotted  with a red solid line.}
\label{fig_slope_hist}
\end{figure}

As mentioned in the introduction we want to use the abundance trends ([X/H]) with T$_{c}$ to select the solar sibling candidates.
For this purpose for each individual star from our sample we made a plot of [X/H] versus T$_{c}$ and then applied a liner regression to
fit the trend. We used the slopes of the linear fits (T$_{c}$ slopes) for further analysis. The 50\% equilibrium
condensation temperatures (T$_{c}$) of the elements are taken from \cite{Lodders-03} which varies between 958 K (for Na) and 1659 K (for Ca and Sc).
The distribution of the T$_{c}$ slopes for the stars in the sample is shown in Fig.~\ref{fig_slope_hist}.

\section{Selection of potential candidates}

Stars that have born in the same cluster are predicted to share similar features, such as chemical abundances and ages. 
Our search is based on the criteria discussed in the literature \citep[e.g.][]{Brown-10, Bobylev-11, Batista-12} and a new criterion based on the T$_{c}$ slopes.

\subsection{Chemical tagging: metallicities and T$_{c}$ slope}

In all the previous studies aimed to track the solar siblings, the authors only focused on the iron abundance, which is usually used as a 
proxy for the overall metallicity of a star. In this study we first constrained our selection to stars with metallicity close to the Sun by
$\pm$ 0.1 dex, i.e., $\mid$[Fe/H]$\mid$ $<$ 0.1 dex. This allowed us to select 339 stars out of 881.

Then in the second step we selected the stars which show T$_{c}$ slopes (in absolute value) and error of the slopes smaller than 0.71$\times 10^{-4}$ dex K$^{-1}$.
This value corresponds to the 0.1 dex difference in [X/H] ($\Delta$[X/H])  when the difference in the T$_{c}$ ($\Delta$T$_{c}$) is 701 K%
\footnote{The whole range of the T$_{c}$ for the elements used in this study is 701 K, i.e., 1659 - 958 K.
}. We finished this step with 101 stars with chemical properties most similar to our Sun. 
We note that the limiting values used in this section are quite conservative, taking into account the uncertainties of the individual element abundances
\citep{Adibekyan-12a, Sousa-11}.

The distribution of the T$_{c}$ slopes of the selected stars are presented in Fig.~\ref{fig_slope_hist}. In the same figure we also overplot 
the expected (simulated) distribution of the T$_{c}$ slopes for stars with solar chemical composition ([X/Fe] = 0 dex) and error of the abundances 
as in \cite{Adibekyan-12a}. One can see that the simulated distribution of the slopes is Gaussian and picked at the zero, while the distribution
for the stars in our sample is skewed towards positive slopes. This is probably due to the fact that age and R$_{mean}$ distributions of our 
sample stars are not symmetric around the solar age and R$_{mean}$ (see Adibekyan et al. 2014a in preparation). It is also interesting to note, that
the bimodality of the T$_{c}$ slope distribution for the full sample is because of the existence of two, thin and thick, disk populations in the 
solar vicinity. The second peak, at larger value of T$_{c}$ slope corresponds to the thick disk stars (see Adibekyan et al. 2014a in preparation).

\begin{table*}
\begin{center}
\caption{The main parameters of HIP97507 and HIP61173.}
\label{table-abundance}
\begin{tabular}{cccccccccccccc}
\hline
\hline
\noalign{\vskip0.01\columnwidth}
\tiny
Star & {[}Fe/H{]}$^{*}$ & \emph{T$_{\mathrm{eff}}$}$^{*}$ & $\log\,g$ $^{*}$  & T$_{c}$ slope & $\pi$ & $\mu$RA  & $\mu$Dec &  RV & U & V & W & Age  \tabularnewline
 & dex & K & dex & dex K$^{-1}$ & \textit{mas} & \textit{mas}/yr & \textit{mas}/yr & \multicolumn{4}{c} { -- km s$^{-1}$ --} & Gyr \tabularnewline
\hline
\noalign{\vskip0.01\columnwidth}
HIP97507 & -0.03$\pm$0.05 & 5662$\pm$62 & 4.44$\pm$0.10  & 5.1 $\pm$ 3.2 (x10$^{-5}$) & 18.56 & -23.81 & -39.37 & -1.3 & -3.1  & -9.2  & 6.7 & 3.5 $\pm$ 3.2\tabularnewline
HIP61173 & 0.06$\pm$0.05 & 5888$\pm$63 & 4.14$\pm$0.10  & -0.4 $\pm$ 4.6 (x10$^{-5}$) & 18.74 & -26.05 & 39.09 & -1.4 & -10.1  & 2.7  & 5.7 & 5.4 $\pm$ 1.2\tabularnewline

\hline 
\end{tabular}
\end{center}
\noindent
Notes: $^{*}$ The errors are the quadratic sum of the precision and systematic errors.
\end{table*}

\subsection{Stellar kinematics and ages}

Numerical simulations conducted by \cite{Brown-10} showed that solar siblings may have approximately the following values of parallax 
($\pi$) and proper motion ($\mu$): $\pi \geq$ 10 \textit{mas}; $\Delta\pi / \pi \leq 0.1$; $\mu \leq$ 6.5\textit{ mas} yr$^{-1}$.
They also quote that stars with these kinematic features may have radial velocities (RV) less than $\sim$10 km s$^{-1}$, in absolute value.
In their simulations they used the observationally established value of (V$_{LSR}$ + V$_{\odot}$)/R$_{\odot}$ constrained by  \cite{McMillan-10} in order to avoid introducing biases 
related to inadequacies in the simulated phase-space distribution of the siblings.

\cite{Bobylev-11} looked for this problem by another perspective. They constrained their search for solar siblings to stars which 
have their magnitude of the total stellar space velocity (henceforth V$_{pec}$ for short) relative to the Sun: V$_{pec}$ = 
(U$^{2}$ + V$^{2}$ + W$^{2}$)$^{1/2} \lesssim$ 8 km s$^{-1}$.  This limiting value was indeed estimated from a typical random 
error, which for each (U,V,W) component is about 22 km s$^{-1}$. Then, the authors analyzed the parameters of their encounters with the solar orbit in the past
in a time interval comparable to the lifetime of stars and selected the best candidates.

Following the criteria introduced by \cite{Brown-10} and also the first criteria of \cite{Bobylev-11}, i.e., V$_{pec}$ $\lesssim$ 8 km s$^{-1}$ 
we did not find any solar sibling candidate within the 101 solar-metallicity stars seleted in the Sect. 3.1.
Indeed, the above mentioned criteria strongly depend on many assumption and are not very strict. Changes in the model parameters would make 
changes also in the final criteria.
Therefore, we choose to slightly extend the kinematic criteria to search stars close to the Sun ($\pi \geq$ 10 \textit{mas} and  $\Delta\pi / \pi \leq 0.1$) 
and to have V$_{pec}$ $<$ 12 km s$^{-1}$. Now, we found two stars, HIP97507 and HIP61173, which have V$_{pec}$ of 11.8 and 11.9 km s$^{-1}$ (see Table 1) 
and very small radial velocity of -1.3 and 1.4 km s$^{-1}$ \citep[][]{Gontcharov-06}, respectively. Taking into account the errors in the velocity components, 
we have decided to consider these stars as plausible solar sibling potential targets. Here we should note, that we did not further compute the past 
encounter parameters with the solar orbit for these stars as it was done in \cite{Bobylev-11}.
The main parameters of these stars are  listed in Table 1.

In the search for solar siblings, one must also look to stellar ages. However, deriving the age of a star is a quite arduous 
task and the errors are often of the order of Gyr. For these two candidate we estimated the age of 3.5$\pm$3.2 (HIP97507) and 5.4$\pm$1.2 (HIP61173)
Gyr by using the PARAM database \citep{daSilva-06} which are comparable with the age of our Sun taking into account the error in age derivation. 
In fact \cite{Casagrande-11} obtained 3.91 and 4.38 Gyr for the age of HIP97507  and 5.16 and 5.9 Gyr using Padova and BASTI isochrones, respectively. 
Our candidate stars have low level of activity, $\log$ $R'_{HK}$ = -4.907$\pm$0.036 (HIP97507) and $\log$ $R'_{HK}$ = -5.119$\pm$0.031 (HIP61173)
which means that the stars are not younger than two Gyr \citep[][]{Pace-13}.
However, as suggested by \cite[][]{Pace-13},  there is no decay of chromospheric activityit after 2 Gyr, and it is not possible to estimate the exact and precise age 
for stars older than the mentioned age.

\begin{figure}
\begin{center}
\begin{tabular}{c}
\includegraphics[angle=270,width=0.9\linewidth]{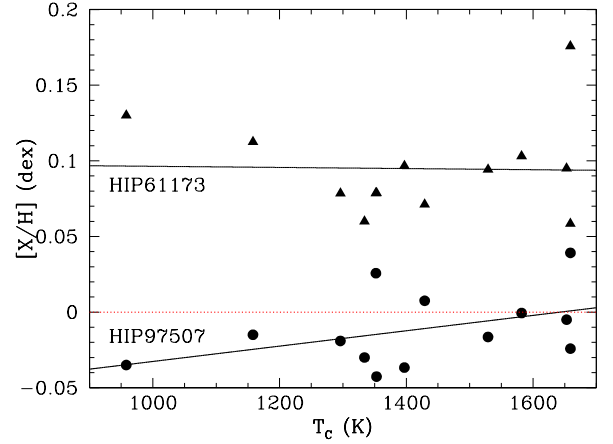}
\end{tabular}
\end{center}
\vspace{-0.6cm}
\caption{[X/H] versus T$_{c}$ for the solar sibling candidates  HIP97507 (black circles) and HIP61173 (black triangles).
The black solid line provides linear fit to the data points. The red dotted line shows the position of the Sun.}
\label{fig_el_tc}
\end{figure}

\section{Results and discussion}

Despite not being the most abundant element, iron is usually used as a proxy of the overall metallicity of a star, 
due to its high number of available spectral lines to measure in solar-type stars. Indeed, the metallicity 
criterion, used in many studies, can be improved by using additional chemical abundances of other elements.

For the first time, this study provides a search for solar siblings in a very homogeneous sample, in terms of 
the metallicity, and chemical abundances in general. We constrained our search for solar siblings among the HARPS FGK dwarfs sample 
\citep{Adibekyan-12a}, based on the trends observed between the chemical abundances of the stars and condensation temperature.
Further applying the kinematics criteria we found two solar sibling candidate, HIP97507 and HIP61173,  which have a peculiar velocity of about 12 km s$^{-1}$.
Our estimation of the age for these stars are very close to the age of our Sun (within the errors).

The [X/H] versus T$_{c}$ for the solar sibling candidates are shown in Fig.~\ref{fig_el_tc}.
As one can see  although HIP61173 shows a practically negligible slope, abundances of all the individual elements are higher 
than 0.05 dex. 
Since this star is a ``solar analog'' the derived stellar parameters and chemical abundances are very 
precise \citep[see][]{Sousa-08, Adibekyan-12a}, which forces us to conclude that HIP61173 does not have solar composition, and hence 
cannot be a solar sibling plausible candidate. 

On the meantime time our paper was submitted,  \cite{Ramirez-14} 
made a detailed analysis of chemical composition of 30 solar sibling candidates listed in the
literature with the goal to select the most probable candidate(s). The authors also showed that several key elements, like Na, Al, V, B, and Y can
be used for a chemical tagging of solar siblings. Following the referee’s suggestion, we derived abundances of several heavy elements including Ba and Y 
for our candidate star (HIP97507): [Cu/H] = 0.02, [Zn/H] = 0.04, [Ba/H] = 0.07 dex,  and [Y/H] = 0.00 dex 
\citep[see][for details on derivation of these elements]{Jonay-10}. These values are quite compatible with the 
composition of the Sun, probably expect Ba. However, several studies showed that the dispersion in Ba abundance within open clusters can be higher than 
0.1 dex \citep[e.g.][]{Jacobson-13, Mishenina-13}. Chemical abundances for the aforementioned elements also suggest a slightly enhanced 
metallicity for HIP61173: [Cu/H] = 0.14, [Zn/H] = 0.07, [Ba/H] = 0.05 dex,  and [Y/H] = 0.00 dex.

Summarizing, we can state that in the sample of 1111 FGK dwarfs from the HARPS GTO subsample \citep{Adibekyan-12a} there 
is only one promising potential candidate: HIP97507 (HD186302).
We note, that the simulations done by \cite{Zwart-09} predict only very few solar siblings within 100 pc from the Sun, and probably with masses
lower than that of the Sun (mostly M dwarfs).
With the new astrometric measurements that will 
be done by Gaia, the knowledge of the stars kinematics of the solar neighborhood will be improved. 
Consequently, the kinematics and age criteria to search for the solar siblings will also be improved.

%
\begin{acknowledgements}

{We thank the referee, Ivan Ramirez, for the very useful comments which helped us to improve the manuscript.
This work was supported by the European Research Council/European Community under the FP7 through Starting Grant agreement 
number 239953. V.Zh.A., S.G.S., and E.D.M. are supported by grants SFRH/BPD/70574/2010, 
SFRH/BPD/47611/2008, and SFRH/BPD/76606/2011 from the FCT (Portugal), respectively.
NCS was supported by FCT through the Investigador FCT contract reference IF/00169/2012
and POPH/FSE (EC) by FEDER funding through the program
 "Programa Operacional de Factores de Competitividade - COMPETE. }
\end{acknowledgements}

\bibliography{refbib}

\end{document}